\documentclass[aps,prb,superscriptaddress,twocolumn]{revtex4-1}
\usepackage[pdftex]{color,graphicx}
\usepackage{amssymb,amsmath}

\begin{document}
\title{A fully integrated high-Q Whispering-Gallery Wedge Resonator}

\author{Fernando Ramiro-Manzano}
\affiliation{Nanoscience Laboratory, Dept. Physics, University of Trento, Via Sommarive 14, I-38050 Trento, Italy}
\author{Nikola Prtljaga}
\affiliation{Nanoscience Laboratory, Dept. Physics, University of Trento, Via Sommarive 14, I-38050 Trento, Italy}
\author{Lorenzo Pavesi}
\affiliation{Nanoscience Laboratory, Dept. Physics, University of Trento, Via Sommarive 14, I-38050 Trento, Italy}
\author{Georg Pucker}
\affiliation{Advanced Photonics \& Photovoltaics Unit, Fondazione Bruno Kessler,  via Sommarive 18, I-38123 Trento, Italy}
\author{Mher Ghulinyan}
\affiliation{Advanced Photonics \& Photovoltaics Unit, Fondazione Bruno Kessler,  via Sommarive 18, I-38123 Trento, Italy}

%

\maketitle \noindent \textbf{Microresonator devices which posses ultra-high quality factors are essential for fundamental investigations and applications. Microsphere and microtoroid resonators support remarkably high Q's at optical frequencies, while planarity constrains preclude their integration into functional lightwave circuits. Conventional semiconductor processing can also be used to realize ultra-high-Q's with planar wedge-resonators. Still, their full integration with side-coupled dielectric waveguides remains an issue. Here we show the full monolithic integration of a wedge-resonator/waveguide vertically-coupled system on a silicon chip. In this approach the cavity and the waveguide lay in different planes. This permits to realize the shallow-angle wedge while the waveguide remains intact, allowing therefore to engineer a coupling of arbitrary strength between these two. The precise size-control and the robustness against post-processing operation due to its monolithic integration makes this system a prominent platform for industrial-scale integration of ultra-high-Q devices into planar lightwave chips.}

Confining photons in a tiny dielectric volume of an ultra-high-Q (UHQ) cavity  increases dramatically light-matter interactions. For this reason, high-Q resonators have been used for a number of fundamental investigations \cite{Aoki,forces,comb,coptomech,particle} and applications \cite{cuttedge,lasers1,lasers2}. In particular, UHQ resonators have been employed for a vast spectrum of studies in quantum photonics \cite{Aoki,Vernooy}, lasing \cite{lasers1,lasers2}, nonlinear optics \cite{comb,opo-hydex,gaeta}, telecommunications \cite{Murugan,flipflop}, and sensing \cite{sensing1,sensing2}.

Engineering of UHQ's in whispering-gallery type resonators has become an enticing objective. One of the challenges to reach such $Q$ values relies on reducing the light scattering at the interface between the resonator and the environment. This way UHQ's were achieved by reflowing the resonator material, as it has been demonstrated for silica-based microspheres\cite{sphere} and microtoroids\cite{toroid}. The non-planarity, however, is the drawback which limits their integration with dielectric waveguides and, consequently, into planar photonic circuits. Another approach for achieving UHQ's is to engineer the geometry of the planar microdisk resonators by realizing a shallow-angle wedge at its rim\cite{1stwedge,2ndwedge}. As a result, the fundamental modes of the wedge-resonator are pushed far away from the scattering edge of the device and, hence, suffer less from surface-induced losses. Very recently, record UHQ's of $\sim10^9$ have been reported for large diameter (5~mm) and relatively thick (10~$\mu$m) silica wedge resonators\cite{leearxiv}.

\begin{figure}[t!]
\centering\includegraphics[width=6cm]{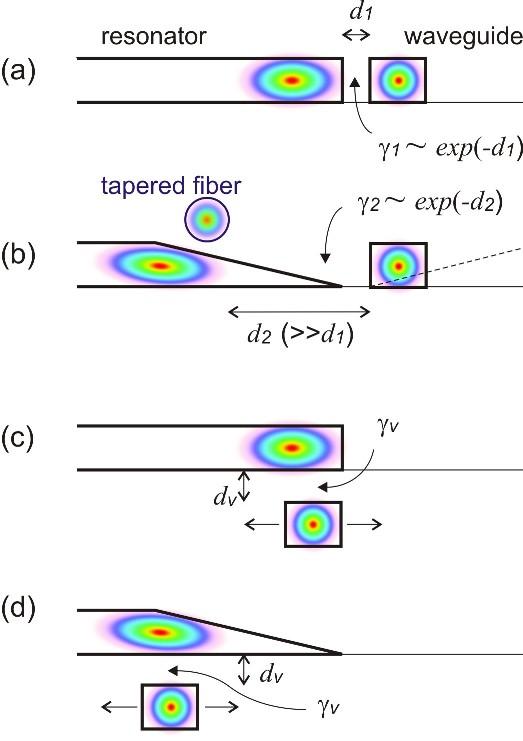}
\caption{
\textbf{Evanescent field coupling schemes in integrated photonics.} (a) A nanometric gap, $d_1$, between a microdisk resonator and a dielectric waveguide defines a side-coupling rate of $\gamma_1\sim\exp(-d_1)$. (b) Such coupling becomes increasingly inefficient when the pre-defined cavity/waveguide distance, $d_2$, increases during the formation of the shallow-angle wedge. In a realistic fabrication process also the waveguide retracts forming a wedge-shape (the dashed inclined line), thus, quenching completely the coupling. Because of this, typically, an off-chip technique (tapered-fiber) is used. (c) The vertical coupling allows for both engineering arbitrarily the evanescent field coupling by controlling the vertical gap, $d_v$, and aligning horizontally the waveguide to the resonator mode. (d) More importantly, this configuration enables the integrated photonics technology to access UHQ wedge resonators with buried dielectric waveguides. The vertical coupling scheme, in fact, permits to realize independently the wedge resonator, maintaining an appropriate coupling rate, $\gamma_v\sim\exp(-d_v)$. It allows for a direct access of devices through integrated waveguides and has the flexibility to engineer free-standing devices\cite{i3e-ptl}. } \label{scheme}
\end{figure}

\begin{figure*}[t!]
\centering\includegraphics[width=15cm]{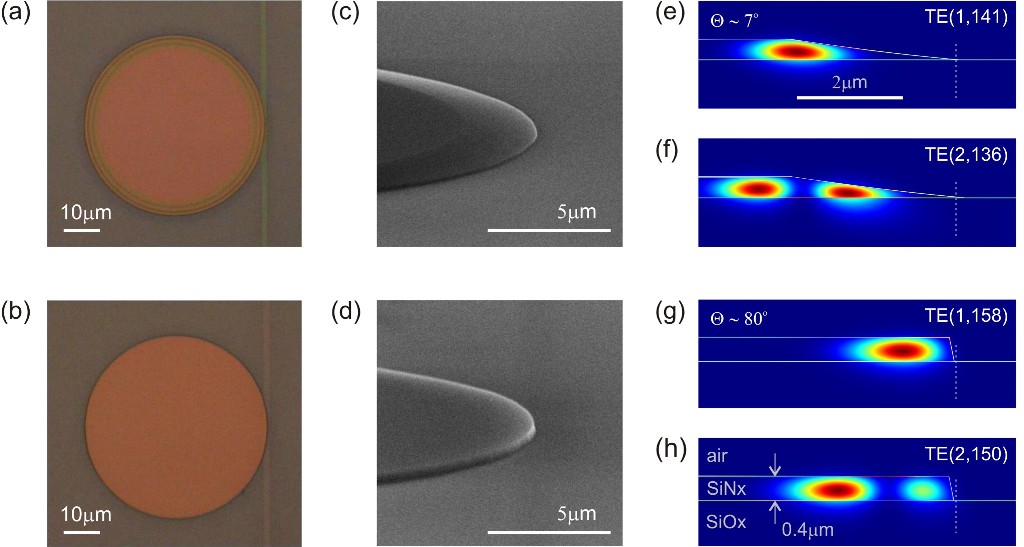}
\caption{
\textbf{Fully integrated wedge  and conventional microdisk resonators with vertical coupling to bus waveguides on a silicon chip.} (a,b) Top-view optical images of the resonators. (a) Concentric Newton's rings of different color at the edge of the wedge resonator indicate to a gradually decreasing layer thickness towards the outer rim. (b) A homogeneous color is observed across the entire disk surface of the dry-etched resonator. Panels (c) and (d) show the bird's-eye-view SEM images of the wet- and dry-etched devices, respectively. The calculated intensities of first and second-order radial modes for both (e,f) the wedge and (g,h) the disk resonators around the wavelength of 1575~nm are shown. These panels illustrate how the wedge resonator modes are pushed away from the device external periphery into an $\approx2.2~\mu$m smaller effective radius trajectory due to the shallow-angled ($\theta\sim7^\circ$) confining wedge.} \label{sem}
\end{figure*}

The planarity of wedge resonators, their CMOS-compatibility and accurate control of the processing entail an important step forward towards a full integration of UHQ devices into planar lightwave circuits. However, a last quest in this direction -- \emph{the integration with on-chip dielectric waveguides} -- is still open. The mode retraction from the cavity rim precludes intrinsically the realization of side-coupled waveguides in the close proximity of the resonator in order to provide an appropriate mode coupling (Fig.~\ref{scheme}a,b). In this work, we show that this important milestone can be reached by opting for a vertical coupling scheme between the wedge microresonator and a bus-integrated waveguide (Fig.~\ref{scheme}c,d). Thereby, this study focuses on the proof-of-concept demonstration of the all-on-chip complete integration of wedge resonators.

Recently, the feasibility of this technology in realizing free-standing microdisk and spiderweb resonators coupled vertically to integrated waveguides through an air gap has been demonstrated\cite{i3e-ptl}. The resonator-waveguide vertical coupling does not require expensive lithographic techniques for the gap definition at a desired precision and allows for an independent choice of materials and thicknesses for optical components. Such advantages can be of great utility in a number of applications to high-speed integrated photonics as well as to cavity optomechanics\cite{coptomech}, requiring high ($\sim10^4$) or ultra-high Q's, respectively. Owing to the separation of the resonator and the waveguide into different planes, our approach enables, on one side, to realize the waveguide and the shallow-angle wedge resonator in different technological steps, and on the other side, to define arbitrary and independently the vertical coupling gap size.

We realized 400~nm-thick and  $50~\mu$m-diameter silicon nitride (SiN$_x$) wedge resonators vertically coupled to silicon oxynitride (SiO$_x$N$_y$) waveguides using standard silicon microfabrication tools (Fig.~\ref{sem}a,c, see also Methods). In order to prove the suppression of surface-induced losses in a fully integrated device, we also realized devices identical to wedge resonators with the only difference of using a conventional dry ion-etching step for the resonator definition (Fig.~\ref{sem}b,d). Spectroscopic characterization and numerical mode analysis were thus performed for both types of devices.

\begin{figure*}[t!]
\centering\includegraphics[width=16cm]{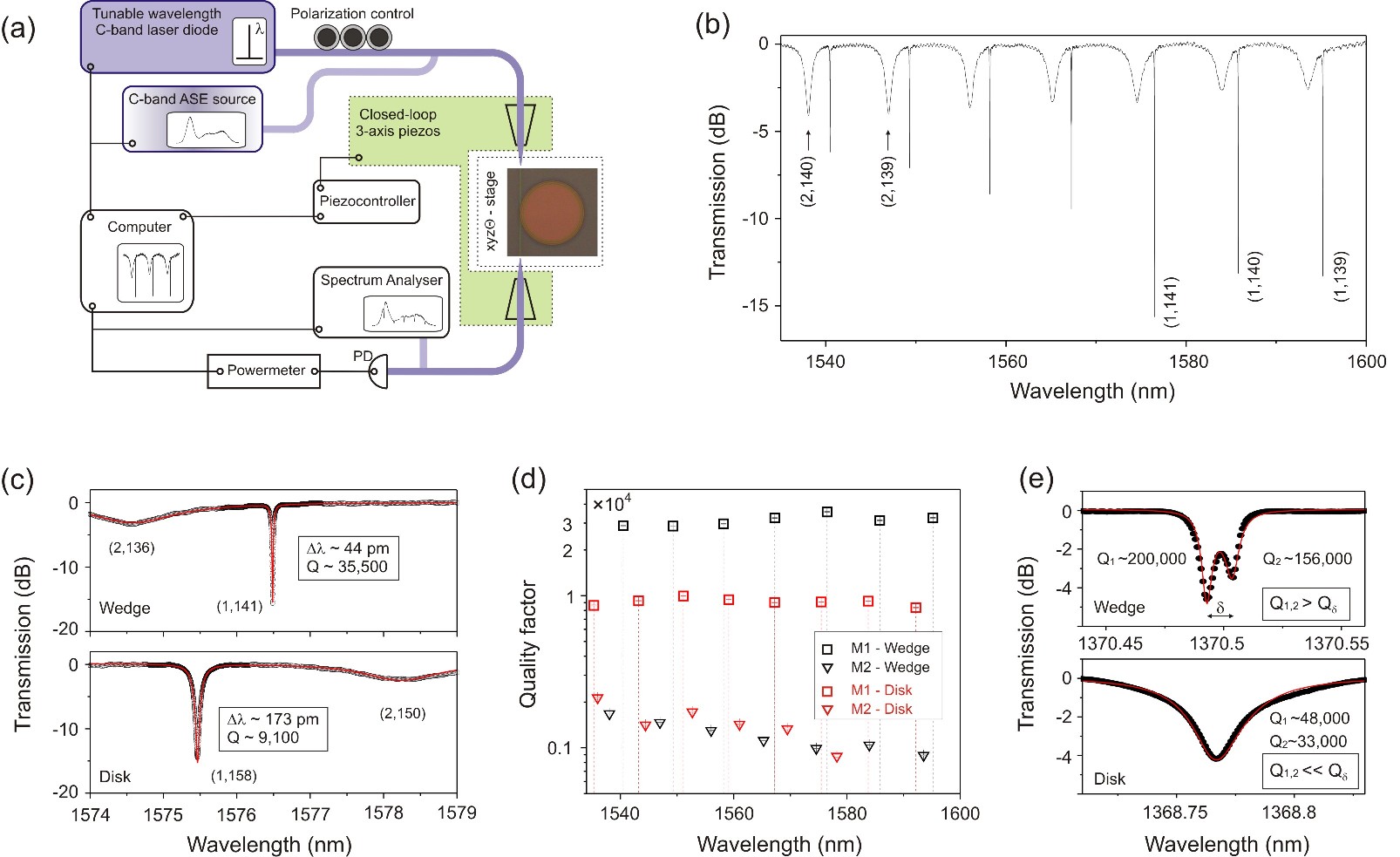}
\caption{\textbf{Optical characterization of wedge resonators.} (a) The optical spectroscopic setup used for waveguide transmission measurements. (b) The measured broad range spectrum of the wedge resonator shows a series of 1st and 2nd order radial family modes of TE-polarization. Some of the these are labeled as ($p,m$) according to their radial $p$ and azimuthal $m$ mode numbers. (c) The high-resolution spectra taken around a wavelength of 1576~nm are shown for the wedge (top panel) and the microdisk (bottom panel) resonators. In both cases the 1st radial family mode is critically coupled to the bus waveguide and has a --15~dB transmission. Lorentzian fits to the resonances reveals a an almost fourfold increase in the 1st order mode's $Q$ for the wedge resonator. (d) The mode Q-factors of both the wedge and dry-etched resonators are compared in a 65~nm range around the wavelength of 1570~nm, showing a three- to fourfold difference for the first order radial families and almost no difference for the second order ones. (e) At a shorter wavelength the wedge resonator's fundamental mode is split into a doublet due to degeneracy lift-off between clockwise and counter-clockwise propagating modes. A reduction in scattering loss of the wedge device and larger transparency of the SiN$_x$ material provide intrinsic Q's higher than the scattering Q$_s$, allowing for doublet observation. Meanwhile, the stronger scattering in the dry-etched device results into a lossy non-split lineshape. } \label{spectra}
\end{figure*}

The wedge resonator was formed during the transfer of the circular photoresist pattern into the SiN$_x$ layer. Due to the adhesion properties of the photoresist a $\theta=7^{\circ}$ sharp angle wedge is formed by the end of etching. Numerical calculations show that the fundamental mode is retracted to a 2.2~$\mu$m smaller effective radius, $r_{eff}$, with respect to the dry-etched (microdisk, hereafter) resonator (Fig.~\ref{sem}e,g).

The devices were characterized in typical waveguide transmission experiments in a broad near-infrared wavelength range between 1350~nm and 1600~nm (Fig.~\ref{spectra}a). Figure ~\ref{spectra}b shows a series of sharp and broad resonances corresponding to first- (fundamental) and second-order radial mode families of the wedge resonator. A blow-up of the spectrum around a fundamental mode with $-15$~dB of transmission suppression is shown in the top panel of Fig.~\ref{spectra}c. By accounting for the critical coupling of this mode to the waveguide, an intrinsic Q$_i\approx7.6\times10^4$ is found from a Lorentzian fit. Interestingly, the fundamental mode of the disk device at the very similar wavelength shows an intrinsic Q$_i$ of only $1.8\times10^4$ (bottom panel in Fig.~\ref{spectra}c). A similar three to fourfold difference in quality factors of the fundamental modes in the wedge and the disk resonators is observed over a broad spectral range (Fig.~\ref{spectra}d). Contrary, the second order mode families show much lower and close $Q$'s over the analyzed spectral range.

At a shorter wavelength ($\sim1370.5$~nm) we observe that the transmission curve at the wedge resonator's undercoupled fundamental mode $M_w=166$ is split into a doublet of symmetric and antisymmetric modes due to scattering. The individual resonances have $Q_{1}=2\times10^5$ and $Q_{2}=1.56\times10^5$, respectively  (Fig.~\ref{spectra}e, top panel).
A fit to the spectrum, using the model for a coherent sum of Lorentzian lineshapes\cite{borselliOE}, shows that the Q's of the individual modes are narrower than the scattering-related $Q_s=\omega_0/\gamma_s$ ($\omega_0$ is the uncoupled mode frequency and $\gamma_s$ is the scattering rate). This indicates to the formation of largely separated standing wave modes. Contrary, no doublet was observed for the disk's (similarly undercoupled) fundamental mode at $\lambda\approx$1368.7~nm ($M_d=186$) (Fig.~\ref{spectra}e, bottom panel). In this case, an important linewidth broadening ($Q_{1,2}\sim3.5\times10^4$) and no observable doublet splitting ($Q_{1,2}<<Q_s$) were found from the sum-Lorentzian fit.

\begin{figure}[t!]
\centering\includegraphics[width=8.5cm]{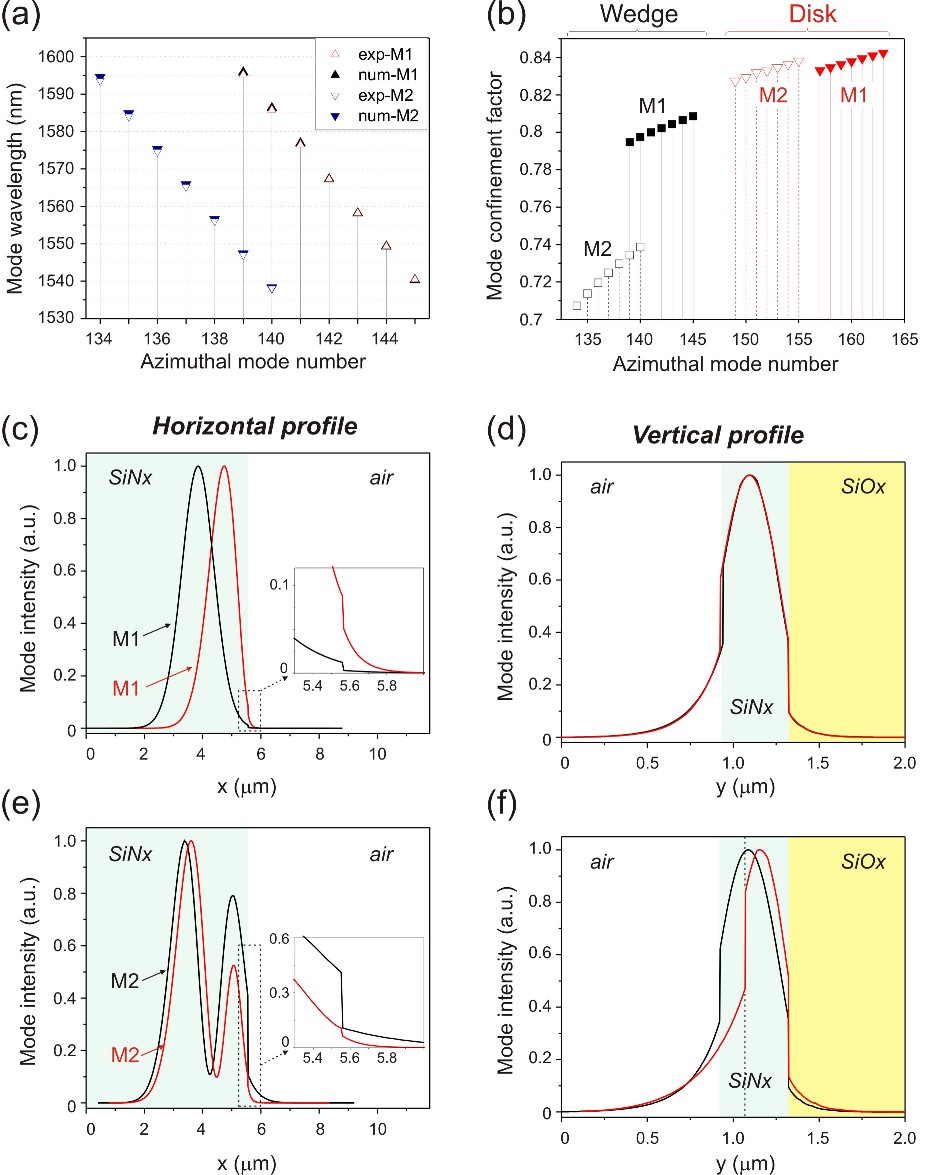}
\caption{\textbf{Numerical analysis of mode confinement and intensity profiles.} (a) Numerically calculated peak wavelengths of the first (M1) and the second (M2) order radial modes of the wedge resonator show an excellent agreement with the experimentally measured ones. (b) The calculated mode confinement factors are reported both for the wedge and the dry-etched resonators. (c) The horizontal and (d) the vertical profiles (through field maxima) of the first order radial modes of two types of resonators around $\lambda=1576$~nm. In particular, the mode maximum in the wedge resonator is situated at a $\sim1\mu$m smaller radius with respect to the dry-etched case, therefore, is more isolated form the physical edge of the device.  The vertical profiles show an identical confinement in both cases. Panels (e) and (f) show the situation for the second order radial modes. In this case, the mode M2 in the wedge resonator is closely situated near the physical edges of the resonator. Insets in (c) and (e) show a close-up of the intensity profiles at the resonator/air interface. } \label{fem}
\end{figure}

In Fig.~\ref{fem}a we compare the measured resonance positions of two radial families of the wedge resonator to the numerically calculated ones. The observed sub-nanometer-precision matching of mode wavelengths permits us to use the numerical model for further analysis of the modal characteristics of the devices. The total intrinsic loss of a whispering-gallery resonator is given as $1/Q_i=1/Q_r+1/Q_m+1/Q_s$, where $Q_r^{-1}$ is the radiative loss, $Q_m^{-1}$ is the modal loss related to the material absorption and $Q_s^{-1}$ is the loss contribution accounting for both the surface scattering and surface absorption. The radiative Q's can be estimated using the ray-optics approach\cite{Levi} and for $50~\mu$m-diameter devices are beyond $10^{19}$, having therefore an increasingly negligible contribution to the overall $Q_i$ for both types of resonators. The quality factor, related to the material and modal loss is $Q_m=\lambda(\Gamma\alpha\Lambda r_{eff})^{-1}$, where $\alpha$ is the bulk material's absorption coefficient, $\Lambda$ is the free-spectral range of the cavity modes, $\lambda$ is the wavelength and $\Gamma$ is the confinement factor of the mode. Figure \ref{fem}b shows that the first radial family modes of the wedge resonator have a confinement of $\sim80\%$ while the $\Gamma$'s in the disk are slightly ($\sim4\%$) larger. Considering that the free-spectral ranges are of about 9.1~nm and 8~nm in the wedge and the disk resonators, respectively, the estimated $Q_m$'s for both devices are very similar.

The last contribution $Q_s^{-1}$ is taking into account the sidewall effects in terms of mode's interaction with the resonators external rim by scattering due to roughness and absorption by specimen on the surface. The horizontal profiles across the fundamental modes brightest spot (Fig.~\ref{fem}c) show that the mode ($M_w=141$) in the wedge resonator is retracted from the etched interface by almost 1~$\mu$m with respect to the disk's case ($M_d=158$) and, consequently, is better isolated from scattering and absorption. On the other hand, the vertical profiles of these modes coincide (Fig.~\ref{fem}d), suggesting that $Q_s$--contribution from planar interfaces is nearly identical for both cases.

This analysis shows that under the conditions of negligible radiative and very similar material losses the main contribution to the intrinsic $Q_i$ comes from the surface-related $Q_s$.
In view of the experimental results (Fig.~\ref{spectra}d), the observed differences in $Q$'s confirm the effective suppression of surface-induced losses in wet-etched devices\cite{1stwedge,2ndwedge,leearxiv}. This conclusions are further supported by the analysis of horizontal (Fig.~\ref{fem}e) and vertical (Fig.~\ref{fem}f) profiles of the second-order mode families. In this case, the modes in the wedge resonator have a significant overlap with the interfaces as compared to the disk's case, hence, larger surface-related losses are expected to significantly limit the $Q$ (see the experimental M2 data-sets in Fig.~\ref{spectra}d).

In conclusion, these results show the feasibility of using conventional silicon microfabrication tools for the realization of planar high-Q wedge resonators monolithically integrated with vertically coupled dielectric waveguides. The integration of waveguides simplifies significantly the access to the resonator, avoiding thus the use of complicated tapered-fiber free-space coupling schemes and guaranteeing stable operation of the device. By opting for a vertical evanescent coupling scheme, several critical issues are resolved; (i) the resonator and the waveguide are fabricated in different fabrication steps without influencing one another, (ii) the wedge angle of the resonator can be controlled without interfering with underlaying waveguiding components, (iii) the waveguide can be freely aligned to the mode position in the resonator, and (iv) the vertical gap between the resonator and the waveguide can be defined for engineering an evanescent coupling of arbitrary strength. The results of this study will boost an intensive research towards the complete integration of fully functional ultra-high quality factor planar resonators into planar photonic circuits. For this ultimate scope, all necessary ingredients, such as the transparent silicon-based materials\cite{leearxiv} as well as an easily accessible technology to realize integrated free-standing devices\cite{i3e-ptl} are already available.\\

\textbf{Methods}\\
\small{
\textbf{Sample fabrication} \\
Monolithically integrated WGM devices vertically coupled to integrated bus waveguides were fabricated starting from the growth of 2.5~$\mu$m-thick thermal oxide on top of crystalline Si wafers. Next,  a 300~nm-thick silicon oxynitride layer was deposited by using plasma-enhanced CVD technique and the strip waveguides defined through standard lithography and reactive ion etching (RIE). These were cladded by a borophosphosilicate silica glass (BPSG) and reflowed for an accurate planarization (96\% degree of planarization). Afterwards, the cladding height was decreased using a RIE in order to define the vertical coupling gap (730~nm's for a critical coupling at 1.55~$\mu$m wavelength). Next, 400~nm's of silicon nitride was deposited using PECVD and patterned lithographically. The resonators were realized by transferring the photoresist to the SiN$_x$ using (i) a wet etching in buffered HF solution to form the wedge resonators and (ii) by dry reactive-ion etching to form microdisk devices (more details in Ref.~[21]).\\

\noindent\textbf{Numerical calculations}\\
Numerical simulations of the resonators have been performed using the COMSOL/FEMLAB's PDE-solver to calculate the fields and frequencies of axisymmetric dielectric resonators. In particular, in order to calculate the resonant frequencies of real devices, the following procedure was used; First, the cross-sectional geometry of the resonator and the real refractive index (found from ellipsometric measurements, $n=1.959$ for SiN$_x$ at $1.55~\mu$m) were set. The resonator's radius was adjusted each time within few nm's around 25~$\mu$m, to allow for processing-induced variations, and the numerical code was run. This procedure was repeated until the calculated mode wavelength coincides with the measured one. Afterwards, the radius was fixed and the rest of the mode wavelengths were obtained changing the azimuthal mode number $M$ only. Such calculated values match to the experimental ones with a sub-nm precision (Fig.~\ref{fem}a).\\

\noindent\textbf{Spectroscopic characterization}\\
The devices were characterized in waveguide transmission experiments exploiting the setup described in Fig.~\ref{spectra}a. The monolithic integration of waveguides permits to use a simple, typical for waveguides characterization experimental setup. A near-infrared tunable laser source was butt-coupled to the bus waveguide, the signal polarization was controlled at the waveguide input and the transmitted power was recorded in a photodiode. For a fast monitoring of spectra and pre-alignment of input fibers a broadband ASE source was also used and the signal was monitored in a optical spectrum analyzer. All the setup was remote controlled by a computer.\\

\textbf{Acknowledgements}\\
This work has been supported partially through NAoMI FUPAT project. One of the authors (F.R.M.) thanks the project APPCOPTOR financed through the FP7 EU Marie Curie fellowship. M. G. and G. P. acknowledge the support of the staff of the Microfabrication Laboratory of FBK during sample fabrication.\\

\textbf{Author contributions}\\
F.R.M performed the experiments and analyzed the data with N.P. All authors contributed to the discussions of results. M.G. conceived the experiments and wrote the paper in collaboration with all authors. Correspondence and requests for materials should be addressed to M.G. (e-mail: ghulinyan@fbk.eu). \\


\end{document}